\documentclass[twocolumn,showpacs,preprintnumbers,amsmath,amssymb,superscriptaddress]{revtex4}
\usepackage{epsfig} 
\usepackage{graphicx}
\usepackage{dcolumn}
\usepackage{bm}

\newcommand{\bee}{\begin{equation}}
\newcommand{\ee}{\end{equation}}
\newcommand{\beea}{\begin{eqnarray}}
\newcommand{\eea}{\end{eqnarray}}

\begin{document}


\title{
Renormalisation group evolution for the $\Delta S = 1$ effective Hamiltonian
with $N_f=2+1$ 
}
\author{David H.~Adams}
\email{dadams@phya.snu.ac.kr}
\affiliation{
  Department of Physics and Astronomy,
  Seoul National University,
  Seoul, 151-747,
  South Korea
}
\author{Weonjong Lee}
\email{wlee@phya.snu.ac.kr}
\affiliation{
  Frontier Physics Research Division and
  Center for Theoretical Physics,
  Department of Physics and Astronomy,
  Seoul National University,
  Seoul, 151-747,
  South Korea
}
\date{\today}
\begin{abstract}
We discuss the renormalisation group (RG) evolution for the 
$\Delta S = 1$ operators in unquenched QCD with $N_f = 3$ 
($m_u=m_d=m_s$) or, more generally, $N_f = 2+1$
($m_u=m_d \ne m_s$) flavors.
In particular, we focus on the specific problem of how to treat the
singularities which show up only for $N_f=3$ or $N_f = 2+1$ in the
original solution of Buras {\it et al.} for the RG evolution matrix 
at next-to-leading order.
On top of Buras {\it et al.}'s original treatment, we use a new method 
of analytic continuation to obtain the correct solution in this case.
It is free of singularities and can therefore be used in numerical
analysis of data sets calculated in lattice QCD.
\end{abstract}
\pacs{11.15.Ha, 12.38.Gc, 12.38.Aw}
\maketitle
\section{Introduction}
In the standard model, the direct CP violation parameter
$\epsilon'/\epsilon$ and weak decays of hadrons are described by a low
energy effective Hamiltonian which can be obtained by decoupling the
heavy quarks and gauge bosons.
In this paper, we consider the $\Delta S =1$ effective Hamiltonian
\cite{ref:buras:npb408,ref:buchalla:rmp68,ref:buras:npb370,
ref:ciuchini:plb301,ref:ciuchini:npb415} 
which governs $\epsilon'/\epsilon$ and the $\Delta I =1/2$ rule.
The Hamiltonian is composed of hadronic matrix elements of four
fermion operators with Wilson coefficients which are known up to 
next-to-leading order (NLO) in perturbation theory
\cite{ref:buchalla:rmp68,ref:nnlo}.
Since the energy scale in kaon decays is about 500 MeV, the hadronic
matrix elements are dominated by the strong interaction, QCD.
Hence we must introduce a non-perturbative tool such as lattice gauge
theory in order to calculate them
\cite{ref:RBC:1,ref:CP-PACS:1,ref:wlee:0,ref:wlee:1}.
Once we calculate the hadronic matrix elements on the lattice, we need
to convert them into the corresponding quantities defined in a
continuum renormalization scheme such as naive dimensional 
regularization (NDR); this is often called ``matching''.
When we match the lattice results to the continuum, we must introduce
a matching scale $q^*$ \cite{ref:hornbostel:1}.
A typical choice of $q^*$ lies in the range from $1/a$ to $\pi/a$
\cite{ref:wlee:2}.
Once we match the lattice results to those in the continuum NDR scheme
at the $q^*$ scale, we have two options to combine the Wilson
coefficients with the continuum results.
One is to run the hadronic matrix elements from $q^*$ down to $m_c$
($N_f=3$) and the other is to run the Wilson coefficients from $m_c$
up to $q^*$.
To do this we use the RG evolution equation at NLO, which is explained
in great detail in
\cite{ref:buras:npb408,ref:buchalla:rmp68,ref:buras:npb370}.
In this paper we focus on the RG evolution equation and its solution
at NLO.
For three sea quark flavors ($N_f=3$ or $N_f=2+1$), singularities
arise in the NLO solution given in \cite{ref:buras:npb408}, even though
the full RG evolution matrix is finite.
This makes it impossible to calculate the RG evolution matrix
numerically in this case, which is an essential step for the lattice
evaluation of $\epsilon'/\epsilon$ and kaon decay amplitudes.
Two unsatisfactory approaches to dealing with this problem have been
attempted previously in the literature.
In Ref.~\cite{ref:RBC:1,ref:RBC:2}, the Wilson coefficients for
$N_f=3$ were combined with the hadronic matrix elements calculated
using the RG evolution matrix with $N_f=0$ (quenched QCD).
In Ref.~\cite{ref:CP-PACS:1,ref:noaki:1}, the singularities were
removed artificially by putting in an arbitrary cut-off of $\approx
1000$ by hand in the calculation of the $N_f=3$ RG evolution matrix
({\em i.e.} singular matrix elements were replaced by finite ones with
this value).
In this paper we provide the correct solution for this problem.
Singularities do not arise, and it can therefore be used for numerical
calculations.
The applicability of our result is broad enough that any lattice
calculation regardless of fermion discretization can take advantage
of it.
In fact, the results of this paper are already being used for the
ongoing data analysis of the staggered $\epsilon'/\epsilon$ project
\cite{ref:wlee:0,ref:wlee:1}.

Unquenched lattice simulations with $N_f=2+1$ sea quark flavors are
currently underway with a variety of fermion discretizations: AsqTad
staggered fermions, HYP staggered fermions, Wilson clover fermions,
twisted mass Wilson fermions, domain wall fermions, and overlap
fermions (see, {\em e.g.}, \cite{ref:wlee:3} and references therein).
These simulations are currently focusing on decay constants ($f_\pi$
and $f_K$), hadron spectrum, the indirect CP violation parameter
$B_K$, kaon semileptonic form factors, and kaon distribution
amplitudes \cite{ref:wlee:3}.
In the coming future, the simulations will be extended to address kaon
physics such as $\epsilon'/\epsilon$ and the $\Delta I = 1/2$ rule.
The results of this paper will be needed in connection with this.

This paper is organized as follows.
In Sec.~\ref{sec:review}, we review the RG evolution results
originally presented in \cite{ref:buras:npb408} and raise the
serious problems for $N_f=3$.
In Sec.~\ref{sec:QCD:part}, we provide the correct solution to the QCD
part of the RG evolution equation, in which singularities do not
arise.
In Sec.~\ref{sec:QED:part}, we go on to present the correct solution
to the QED part of the RG evolution equation, in which there are also
no singularities.
We close with some conclusions.
\section{Review of RG evolution}
\label{sec:review}
The $\Delta S = 1$ effective Hamiltonian for non-leptonic decays may
be written in general as
\begin{eqnarray}
{\cal H}_{eff} = \frac{G_F}{ \sqrt{2} } \sum_{i} C_i(\mu) Q_i(\mu)
\equiv \frac{G_F}{ \sqrt{2} } \vec{Q}^T(\mu) \cdot \vec{C}(\mu)
\end{eqnarray}
where the index $i$ runs over a basis for the contributing operators; 
in our example this is the basis $Q_1$, $Q_2$, $\cdots$, $Q_{10}$ of
of Buras {\it et al.} defined in Section 2 of 
Ref.\cite{ref:buras:npb408}.    
The renormalization group equation for $\vec{C}(\mu)$ is
\begin{eqnarray}
\left[ \mu \frac{\partial}{\partial \mu} 
+ \beta(g) \frac{\partial}{\partial g} \right] \vec{C} = 
\gamma^T(g,\alpha) \vec{C}
\label{eq:rg}
\end{eqnarray}
where $\beta(g)$ is the QCD beta function:
\begin{equation}
\beta(g) = 
- \beta_0 \frac{g^3}{ 16\pi^2 }
- \beta_1 \frac{g^5}{ (16\pi^2)^2 }
- \beta_{1e} \frac{e^2 g^3}{ (16\pi^2)^2 }
\end{equation}
with
\begin{equation}
\beta_0 = 11 - \frac{2}{3} f 
\qquad
\beta_1 = 102 - \frac{38}{3} f
\qquad
\beta_{1e} = - \frac{8}{9} \left( u + \frac{d}{4} \right)
\end{equation}
and $f=u+d$ denoting the number of active flavors, $u$ and $d$ being
the number of $u$-type and $d$-type flavors respectively.
According to Buras {\em et al.}, the the contribution from the
$\beta_{1e}$ term is negligible and they dropped it in Ref.'s
\cite{ref:buras:npb408,ref:buras:npb370}. We will include it here for
the sake of completeness.
The $\gamma(g,\alpha)$ matrix is the full $10 \times 10$ anomalous
dimension matrix, which is given in \cite{ref:buras:npb408}.
The solution of the RG equation in Eq.~(\ref{eq:rg}) for the Wilson
coefficient functions is given by
\begin{eqnarray}
\vec{ C } (\mu) = U( \mu, \mu_W, \alpha) \vec{ C } (\mu_W)
\end{eqnarray}
The coefficients at the scale $\mu_W = {\cal O}(M_W)$ can be evaluated
in perturbation theory.
The evolution matrix $U$ then includes the renormalization group
improved perturbative contributions from the scale $\mu_W$ down to
$\mu$.
For $m_1 < m_2$,
\begin{equation}
U ( m_1, m_2, \alpha) \equiv
T_g \exp \left( 
\int^{ g(m_1) }_{ g(m_2) } dg'
\frac{ \gamma^T(g',\alpha) }{ \beta(g') }
\right)
\label{eq:evol:mat}
\end{equation}
where $g$ is the QCD coupling.
Here, $T_g$ denotes ordering in the coupling constant such that the
couplings increase from right to left.
Note that for $g_1 \ne g_2$,
\begin{equation}
[ \gamma(g_1), \gamma(g_2) ] \ne 0
\end{equation}
The evaluation of the amputated Green functions with insertion of
the operators $\vec{Q}$ gives the following relation
\begin{eqnarray}
\langle \vec{Q} \rangle^{(0)}
= Z^{-2}_q Z \langle \vec{Q} \rangle
\end{eqnarray}
where $\langle \vec{Q} \rangle^{(0)}$ and $\langle \vec{Q} \rangle$
denote the unrenormalized and renormalized Green functions,
respectively.
$Z_q$ is the quark field renormalization constant and $Z$ is the
renormalization constant matrix of the operators $\vec{Q}$.
The anomalous dimension matrices are defined by
\begin{equation}
\gamma (g, \alpha) = Z^{-1} \frac{d}{d \ln\mu} Z
\end{equation}
which includes QCD and QED contributions.
For the case at hand, $\gamma( g, \alpha)$ can be expanded in the
following way (with $\alpha = \frac{e^2}{4\pi}$):
\begin{equation}
\gamma( g, \alpha) = \gamma_s (g^2) +
\frac{\alpha}{4\pi} \Gamma(g^2) + \cdots
\end{equation}
The QCD part of the anomalous dimension, $ \gamma_s $, can be expanded
as (with $\alpha_s = \frac{g^2}{4\pi}$)
\begin{equation}
\gamma_s (g^2) = \frac{ \alpha_s }{ 4 \pi } \gamma^{(0)}_s 
+ \frac{ \alpha_s^2 }{ (4 \pi)^2 } \gamma^{(1)}_s + \cdots
\end{equation}
The QED part of the anomalous dimension, $\Gamma$, can be expanded as
\begin{equation}
\Gamma (g^2) = \gamma^{(0)}_e + 
\frac{\alpha_s}{ 4 \pi } \gamma^{(1)}_{se} + \cdots
\end{equation}
The general RG evolution matrix $U(m_1,m_2,\alpha)$ of
Eq.~(\ref{eq:evol:mat}) may then be decomposed as follows
\begin{eqnarray}
U ( m_1, m_2, \alpha) &=& U( m_1, m_2) + 
\frac{\alpha}{ 4 \pi } R( m_1, m_2)
\label{eq:evol:mat:2}
\\
U(m_1,m_2) &=& T_g \exp \left( 
\int^{ g(m_1) }_{ g(m_2) } dg' \frac { \gamma_s^T(g') } { \beta(g') }
\right) 
\label{eq:evol:qcd}
\\
R( m_1, m_2) &=& \int^{ g(m_1) }_{ g(m_2) } 
dg' \frac{ U( m_1, m') \Gamma^T(g') U( m', m_2) }{ \beta(g') }
\nonumber \\
\label{eq:evol:qed}
\end{eqnarray}
Here, $U(m_1,m_2)$ represents the pure QCD evolution and $R(m_1, m_2)$
describes the additional evolution in the presence of the
electromagnetic interaction.
The leading order RG equation describing the QED evolution was first 
discussed in \cite{ref:buchalla:npb337}. 
The QCD evolution matrix which Buras {\em et al.} provided
originally can be expressed up to NLO as
\begin{eqnarray}
U( m_1, m_2) &=& \left( 1 + \frac{ \alpha_s(m_1) }{ 4 \pi } J \right)
U^{(0)}( m_1, m_2)
\nonumber \\
& &  \cdot \left( 1 - \frac{ \alpha_s(m_2) }{ 4 \pi } J \right)
\label{eq:U:qcd}
\end{eqnarray}
where $U^{(0)}(m_1,m_2)$ denotes the evolution matrix in the leading
logarithmic approximation and $J$ summarizes the next-to-leading 
correction to this evolution.
Additional terms proportional to $\alpha_s^2$ in $U(m_1,m_2)$ which do
not come from $U^{(0)}( m_1, m_2)$ should be consistently dropped at
NLO.
Taking $V$ to be a matrix which diagonalizes $\gamma^{(0) T}_s$, we
define the following:
\begin{eqnarray}
\gamma^{(0)}_D &=& V^{-1} \gamma^{(0) T}_s V
\\
G &=& V^{-1} \gamma^{(1) T}_s V 
\label{eq:G:mat}
\end{eqnarray}
where $\gamma^{(0)}_D$ denotes a diagonal matrix whose diagonal
elements are the components of the vector $\vec{\gamma}^{(0)}$.
Then,
\begin{eqnarray}
U^{(0)}( m_1, m_2) = V \left[ \left( \frac{\alpha_s(m_2)}{\alpha_s(m_1)}
\right)^{\vec{a}} \right]_D V^{-1}
\label{eq:U0}
\end{eqnarray}
with
\begin{eqnarray}
\vec{a} &=& \frac{ \vec{\gamma}^{(0)}_s }{ 2 \beta_0 }
\label{eq:a}
\end{eqnarray}
For the matrix $J$, we find 
\begin{eqnarray}
J &=& V S V^{-1}
\label{eq:J}
\end{eqnarray}
where the elements of $S$ are given by
\begin{equation}
S_{ij} = \delta_{ij} \gamma^{(0)}_i \frac{\beta_1}{2 \beta_0^2}
- \frac{G_{ij}}{2 \beta_0 + \gamma^{(0)}_i - \gamma^{(0)}_j }
\label{eq:S:mat}
\end{equation}
where $\gamma^{(0)}_i$ is a component of $\vec{\gamma}^{(0)}$ and
$G_{ij}$ denotes the elements of $G$ in Eq.~(\ref{eq:G:mat}).
This is the result obtained by Buras {\em et al.} 
\cite{ref:buras:npb408,ref:buras:npb370}. 
However, when $f=3(=N_f)$, $i=8$ and $j=7$, the denominator of
Eq.~(\ref{eq:S:mat}) vanishes: $2\beta_0+\gamma^{(0)}_i-\gamma^{(0)}_j
= 0$.
Hence the solution for the RG evolution matrix at NLO has
singularities in this case, despite the fact that the full RG
evolution matrix must be finite.
This is one of the main issues we will address in this paper.
This problem is briefly mentioned in 
\cite{ref:buras:npb408,ref:ciuchini:npb415} without any solution given.
The correct treatment for this case, which is given in the next
section, eradicates the singularities in Eqs.~(\ref{eq:U:qcd}) and
(\ref{eq:S:mat}), and leads to a finite expression for the RG
evolution matrix at NLO.
Next, we turn to the QED part of the evolution matrix $R(m_1,m_2)$
given in Eq.~(\ref{eq:evol:qed}).
We can expand $R(m_1,m_2)$ in powers of $g^2$ as follows:
\begin{eqnarray}
R(m_1,m_2) &=& R^{(0)}(m_1, m_2) + R^{(1)}(m_1, m_2) + \cdots
\end{eqnarray}
where $R^{(i)}$ is of the order of $g^{2i}$.
It turns out to be convenient to introduce the matrix $K(m_1,m_2)$ to
represent $R(m_1, m_2)$ as follows:
\begin{eqnarray}
R(m_1,m_2) &=& - \frac{2\pi}{\beta_0} V K(m_1,m_2) V^{-1}
\\
K(m_1,m_2) &=& K^{(0)}(m_1,m_2) 
\nonumber \\
&+&
\frac{1}{4 \pi} \sum_{i=1}^{3} K^{(1)}_i(m_1,m_2)
\\
R^{(0)}(m_1,m_2) &=& - \frac{2\pi}{\beta_0}
V K^{(0)}(m_1,m_2) V^{-1}
\\
R^{(1)}(m_1,m_2) &=& 
- \frac{2 \pi}{\beta_0} 
\frac{1}{4 \pi} \sum_{i=1}^{3}
V K^{(1)}_i(m_1,m_2) V^{-1}
\end{eqnarray}
The leading order term can be obtained by straightforward integration
\begin{eqnarray}
( K^{(0)} (m_1, m_2) )_{ij} &=& 
\frac{ M^{(0)}_{ij} }{a_i - a_j - 1}
\bigg[ 
\left( \frac{ \alpha_s (m_2) }{ \alpha_s (m_1) } \right)^{ a_j }
\frac{1}{ \alpha_s(m_1) }
\nonumber \\
& & - \left( \frac{ \alpha_s (m_2) }{ \alpha_s (m_1) } \right)^{ a_i }
\frac{1}{ \alpha_s(m_2) }
\bigg]
\label{eq:K0:1}
\end{eqnarray}
where the $a_i$'s are the components of $\vec{a}$ in Eq.~(\ref{eq:a}),
and the  $M^{(0)}$ matrix is given by
\begin{equation}
M^{(0)} = V^{-1} \gamma_e^{(0)T} V
\end{equation}
Similar to the case of $S_{ij}$ in Eq.~(\ref{eq:S:mat}), there is a
singularity in Eq.~(\ref{eq:K0:1}) for the element (7,8) of $\left(
K^{(0)}(m_1,m_2) \right)$ since $a_7 = a_8 + 1$ when $f=3(=N_f)$.
However, the expression in the numerator also vanishes in this case
and so the singularity is removable.
In this case of $a_i = a_j + 1$, direct integration leads to the
following formula \cite{ref:buras:npb408}:
\begin{eqnarray}
\left( K^{(0)}(m_1,m_2) \right)_{ij} &=&
M^{(0)}_{ij} \frac{1}{ \alpha_s(m_1) }
\left( \frac{ \alpha_s(m_2) }{ \alpha_s(m_1) } \right)^{a_j}
\nonumber \\
& \times &
\ln \left( \frac{ \alpha_s(m_1) }{ \alpha_s(m_2) } \right) 
\label{eq:K0:2}
\end{eqnarray}
The next leading corrections to the QED part of the evolution matrix
are represented by $K^{(1)}_i(m_1,m_2)$.
We introduce
\begin{eqnarray}
\Gamma^{(1)} &=& \gamma^{(1) T}_{se} 
- \frac{\beta_1}{\beta_0} \gamma^{(0) T}_e
- \frac{\beta_{1e}}{\beta_0} \gamma^{(0) T}_s
\end{eqnarray}
and 
\begin{equation}
M^{(1)} = V^{-1} ( \Gamma^{(1)} + [ \gamma^{(0) T}_e, J ] ) V
\end{equation}
The matrices $K^{(1)}_i(m_1,m_2)$ are then given as follows:
\begin{widetext}
\begin{eqnarray}
\left( K^{(1)}_1 (m_1, m_2) \right)_{ij} &=&
M^{(1)}_{ij} Q_{ij}
\label{eq:K1-1}
\\
Q_{ij}
&=& 
\left\{ \begin{array}{ll}
\frac { 1 } { a_i - a_j }
\left[ 
\left( \frac{ \alpha_s(m_2) }{ \alpha_s(m_1) } \right)^{a_j}
- \left( \frac{ \alpha_s(m_2) }{ \alpha_s(m_1) } \right)^{a_i}
\right]
&  \mbox{ if $i \ne j$ } \\
\left( \frac{ \alpha_s(m_2) }{ \alpha_s(m_1) } \right)^{a_i}
\ln \left( \frac{ \alpha_s(m_1) }{ \alpha_s(m_2) } \right)
&  \mbox{ if $i = j$ }
	  \end{array} \right.
\end{eqnarray}
\end{widetext}
\begin{eqnarray}
K^{(1)}_2 ( m_1, m_2) &=& -\alpha_s( m_2 ) K^{(0)}( m_1, m_2) S
\label{eq:K1-2} \\
K^{(1)}_3 ( m_1, m_2) &=& \alpha_s( m_1 ) S K^{(0)}( m_1, m_2)
\label{eq:K1-3}
\end{eqnarray}
As one can see in the above equations, all of the $K^{(1)}_i ( m_1,
m_2)$ matrices include $S$ or $J$.
Since $S$ or $J$ is singular for $f=3(=N_f)$, all the $K^{(1)}_i$
matrices are also singular.
However, the full RG evolution matrix $R(m_1,m_2)$ must always be
finite.
The correct treatment of $R(m_1,m_2)$ at NLO in this case, given in
Section \ref{sec:QED:part}, modifies the formulae given in
Eq.~(\ref{eq:K1-1}-\ref{eq:K1-3}) such that all the singularities are
eradicated.

%
%
%
%
%

%
\section{How to handle removable singularities for $N_f=3$ (QCD part)}
\label{sec:QCD:part}
%
%
For $f=3$ (three dynamical flavors), when $i=8$ and $j=7$,
$\beta_0=9$, $\gamma^{(0)}_i=-16$ and $\gamma^{(0)}_j=2$.
Hence, $2 \beta_0 + \gamma^{(0)}_i - \gamma^{(0)}_j = 0$ corresponds
to a pole in $S_{ij}$ of Eq.~(\ref{eq:S:mat}), which does not exists
in the full evolution matrix and so should not appear in the correct
solution at NLO.
To find the latter, we start by rearranging the NLO expression for the
QCD RG evolution matrix given in Eq.~(\ref{eq:U:qcd}) as follows
\begin{eqnarray}
U(m_1,m_2) 
&=& U_0(m_1,m_2)
+ \frac{1}{4 \pi} V A(m_1,m_2) V^{-1}
\label{eq:U:4}
\end{eqnarray}
where
\begin{eqnarray}
V A(m_1,m_2) V^{-1}
&=& 
\alpha_s(m_1) J U_0(m_1,m_2)
\nonumber \\
& & - \alpha_s(m_2) U_0(m_1,m_2) J 
\label{eq:A:1}
\end{eqnarray}
When $S$ and $J$ matrices are non-singular, the $A$ matrix is readily
found from Eqs.~(\ref{eq:U0})-(\ref{eq:S:mat}) and Eq.~(\ref{eq:A:1})
to be given by 
\begin{eqnarray}
A_{ij} &=& S_{ij} \left[ 
\alpha_s(m_1) \left( \frac{ \alpha_s(m_2) }{\alpha_s(m_1) } \right)^{a_j} 
- \alpha_s(m_2) \left( \frac{ \alpha_s(m_2) }{ \alpha_s(m_1) } \right)^{a_i} 
\right]
\nonumber \\
\label{eq:A:2}
\end{eqnarray}
When $S_{ij}$ is singular ({\em i.e.} $i=8$ and $j=7$), this expression
diverges and therefore cannot be used in numerical calculations.
In this case, the correct finite expression for the $A$ matrix is 
\begin{eqnarray}
A_{ij} = \frac{G_{ij}}{2 \beta_0} 
\alpha_s(m_2) \left(
\frac{\alpha_s(m_2)}{\alpha_s(m_1)} \right)^{a_i} \ln \left(
\frac{\alpha_s(m_2)}{\alpha_s(m_1)} \right)
\label{eq:A:3}
\end{eqnarray}
The derivation is as follows.
For $i \ne j$, 
\begin{eqnarray}
S_{ij} = - \frac{ G_{ij} }{ 2 \beta_0 (1 + a_i - a_j) }
\end{eqnarray}
We regularize the singularity that occurs when $a_j = a_i + 1$ ($i=8$
and $j=7$) by introducing an $\epsilon$ shift of $a_j$ so that $a_j =
a_i + 1 + \epsilon$.
Then
\begin{equation}
S_{ij} = \left( \frac{ G_{ij} }{ 2 \beta_0 } \right) 
\frac{1}{\epsilon}
\end{equation}
and $A_{ij}$ is given by
\begin{eqnarray}
A_{ij} &=& \left( \frac{ G_{ij} }{ 2 \beta_0 } \right)
\frac{1}{\epsilon} \alpha_s(m_2)
\left( \frac{ \alpha_s(m_2) } { \alpha_s(m_1) } \right)^{a_i}
\bigg[ \epsilon \ln \left( \frac{ \alpha_s(m_2) } { \alpha_s(m_1) } \right)
\nonumber \\
& & + {\cal O} (\epsilon^2) \bigg]
\end{eqnarray}
In the limit of $\epsilon = 0$, we get the claimed result of
Eq.~(\ref{eq:A:3}).
The expression Eq.~(\ref{eq:A:3}) is finite and can therefore be used
for numerical calculations.
In the appendix we give an alternative derivation of this result,
similar to the previous approach of Buras {\em et al.} in the 
non-singular case \cite{ref:buras:npb408,ref:buras:npb370,ref:buras:rmp52}. 
\section{How to handle removable singularities for $N_f=3$ (QED part)}
\label{sec:QED:part}
Now let us turn to the QED part of the evolution matrix for $f=3(=N_f)$.
Basically, we want to perform the integration in the right-hand side of
Eq.~(\ref{eq:evol:qed}).
The Eq.~(\ref{eq:evol:qed}) contains the QCD evolution matrix $U$.
Since the $U$ matrix is modified due to the removable singularity for
$f=3$ as given in Eqs.~(\ref{eq:U:4}-\ref{eq:A:3}), the QED part, the $R$
matrix, will also change correspondingly.
In the following we provide this modified version of the $R$ matrix.
As we mentioned in the previous section, we can not use the $S$ and
$J$ matrices because they are singular.
Instead, we need to define a new matrix which is finite:
\begin{eqnarray}
H_{ij} &=& S_{ij} ( 1 - \delta_{i,8} \delta_{j,7} )
\end{eqnarray}
Note that the singular part is subtracted away so that the $H$ matrix
is finite.
We now express the $R$ matrix as follows:
\begin{eqnarray}
R( m_1, m_2) &=& - \frac{2\pi}{\beta_0} V \tilde{K}(m_1,m_2) V^{-1}
\\
\tilde{K}(m_1,m_2) &=& K^{(0)}(m_1,m_2) 
\nonumber \\
&+&
\frac{1}{4 \pi} \sum_{i=1}^{4} \tilde{K}^{(1)}_i(m_1,m_2)
\end{eqnarray}
where the leading term $K^{(0)}$ is given in
Eqs.~(\ref{eq:K0:1}-\ref{eq:K0:2}).
The $\tilde{K}^{(1)}_i$ matrices are given as
\begin{eqnarray}
[ \tilde{K}^{(1)}_1 ]_{ij} &=& 
\big[ M^{(2)} + [M^{(0)}, H] \big]_{ij} Q_{ij}
\label{eq:tK1:1}
\\ {}
[ \tilde{K}^{(1)}_2 ]_{ij} 
&=& \alpha_s (m_1) [ H \ K^{(0)} ]_{ij}
\label{eq:tK2:1}
\\ {}
[ \tilde{K}^{(1)}_3 ]_{ij} &=& -\alpha_s (m_2) [ K^{(0)} \ H ]_{ij}
\label{eq:tK3:1}
\\ {}
[ \tilde{K}^{(1)}_4 ]_{ij} &=& 
\delta_{i8} \bigg[ \frac{ G_{87} M^{(0)}_{7j} }{2 \beta_0} \bigg]
[I_1]_{ij} 
+ \delta_{j7} \bigg[ \frac{ M^{(0)}_{i8} G_{87} }{2 \beta_0} \bigg]
[I_2]_{ij}
\nonumber \\
\label{eq:tK4:1}
\end{eqnarray}
where $M^{(2)}$ is
\begin{eqnarray}
M^{(2)} &=& V^{-1} \Gamma^{(1)} V
\end{eqnarray}
A complicated calculation lead to the following expressions for the
$I_1$ and $I_2$ matrices:
\begin{widetext}
\begin{eqnarray}
[ I_1 ]_{ij} &=& \left\{ 
\begin{array}{ll}
  \frac{1}{a_i - a_j} \bigg( \frac{\alpha_s(m_2)}{\alpha_s(m_1)} \bigg)^{a_i}
  \ln \bigg( \frac{\alpha_s(m_1)}{\alpha_s(m_2)} \bigg)
  - \frac{1}{ (a_i - a_j)^2 } 
  \bigg[
    \bigg( \frac{\alpha_s(m_2)}{\alpha_s(m_1)} \bigg)^{a_j}
    - \bigg( \frac{\alpha_s(m_2)}{\alpha_s(m_1)} \bigg)^{a_i}
    \bigg] 
  & \mbox{if $a_i \ne a_j$} \\
  -\frac{1}{2} \bigg( \frac{\alpha_s(m_2)}{\alpha_s(m_1)} \bigg)^{a_j}
  \bigg[ \ln \bigg( \frac{\alpha_s(m_2)}{\alpha_s(m_1)} \bigg) \bigg]^2
  & \mbox{if $a_i = a_j$}
\end{array} \right.
\\ {}
[ I_2 ]_{ij} &=& \left\{ 
\begin{array}{ll}
  \frac{1}{a_i - a_j} \bigg( \frac{\alpha_s(m_2)}{\alpha_s(m_1)} \bigg)^{a_j}
  \ln \bigg( \frac{\alpha_s(m_2)}{\alpha_s(m_1)} \bigg)
  + \frac{1}{ (a_i - a_j)^2 } 
  \bigg[
    \bigg( \frac{\alpha_s(m_2)}{\alpha_s(m_1)} \bigg)^{a_j}
    - \bigg( \frac{\alpha_s(m_2)}{\alpha_s(m_1)} \bigg)^{a_i}
    \bigg] 
  & \mbox{if $a_i \ne a_j$} \\
  -\frac{1}{2} \bigg( \frac{\alpha_s(m_2)}{\alpha_s(m_1)} \bigg)^{a_j}
  \bigg[ \ln \bigg( \frac{\alpha_s(m_2)}{\alpha_s(m_1)} \bigg) \bigg]^2
  & \mbox{if $a_i = a_j$}
\label{eq:I2}
\end{array} \right.
\end{eqnarray}
\end{widetext}
Note that the leading order contribution $K^{(0)}$ is the same as
before.
The only change is localized in the $\tilde{K}^{(1)}_i$ matrices.
In particular, the $\tilde{K}^{(1)}_i$ matrices for ($i=1,2,3$)
corresponds to the $K^{(1)}_i$ matrices once we substitute the $S$
matrix by the $H$ matrix.
The $\tilde{K}^{(1)}_4$ matrix represents the contribution from the
removable singularity of the $S$ matrix.
The key point is that the $\tilde{K}^{(1)}_i$ matrices are finite
and can be used numerically, whereas the $K^{(1)}$ are divergent.
%


%
\section{Conclusion}
The original solution of Buras {\it et al.} for the RG evolution matrix 
at NLO contains removable singularities for $N_f = 3$, which cancel out 
in the proper combination.
However, since the individual terms are singular, it is not possible
to use it in the numerical calculation.
In this paper, we provide the correct solution in which there are no
singularities.
Our results for both the QCD part and QED part of the RG evolution matrix
are finite and can be used for numerical studies with $N_f = 2+1$.
These results are currently being used to analyze the data sets of
the staggered $\epsilon'/\epsilon$ project \cite{ref:stag:e'/e}.
In fact, this work is a part of that project.
\section{Acknowledgement}
Helpful discussion with S.~Sharpe and A.~Soni is acknowledged with
gratitude.
This research is supported by the KOSEF international cooperative
research program (KOSEF grant M60501000018-06A0100-01810), by the
KOSEF grant (R01-2003-000-10229-0), by the KRF grant (C00497), by the
BK21 program of Seoul National University, and by the DOE SciDAC-2
program.
\appendix*
\section{Alternative derivation of the pure QCD evolution matrix at NLO}
In this appendix we give an alternative derivation of the result for the pure 
QCD evolution matrix $U(m_1,m_2)$ at NLO in the singular case 
(Eq.'s (\ref{eq:U:4}) and (\ref{eq:A:3})). 
Since the dependence of $U(m_1,m_2)$ on $m_i$
enters through $\alpha_s(m_i)=g(m_i)^2/4\pi\,$, we will sometimes denote the 
evolution matrix by $U(g_1,g_2)$ when it is convenient. Following the earlier
approach of Buras {\em et al.} \cite{ref:buras:npb370,ref:buras:rmp52}, 
we express the evolution matrix as
\begin{equation}
U(g,g_0)=\Big(1+\frac{g^2}{16\pi^2}J(g)\Big)\,U^{(0)}(g,g_0)\,
\Big(1+\frac{g_0^2}{16\pi^2}J(g_0)\Big)^{-1}
\label{A1}
\end{equation}
and seek to determine $J(g)$ from the differential equation which 
characterizes $U(g,g_0)$:
\begin{equation}
\frac{d}{dg}U(g,g_0)=\frac{\gamma_s^T(g^2)}{\beta(g)}U(g,g_0)
\label{A2}
\end{equation}
Substituting (\ref{A1}) into (\ref{A2}), and introducing $S(g)$ via 
$J(g)=VS(g)V^{-1}$ as in Eq.~(\ref{eq:J}), we find the following differential
equation for the matrix elements of $S(g)$:
\begin{eqnarray}
\beta_0\,g\,S_{ij}'(g)+(2\beta_0+\gamma_i^{(0)}-\gamma_j^{(0)})S_{ij}(g)
\nonumber \\
-\frac{g^2}{16\pi^2}\Big((\frac{\beta_1}{\beta_0}\gamma_D^{(0)}-G+O(g^2))
S(g)\Big)_{ij}
\nonumber \\
=\frac{\beta_1}{\beta_0}\gamma_i^{(0)}\delta_{ij}-G_{ij}+O(g^2)
\label{A4}
\end{eqnarray}
To solve this equation at leading order in $g$ it is necessary to consider
separately the cases where $2\beta_0+\gamma_i^{(0)}-\gamma_j^{(0)}$ is 
non-vanishing and vanishing. In the former case, a consistent solution is
obtained at lowest order by taking $J(g)=J+O(g)$ where $J$ is a constant 
matrix. Then $S(g)=S+O(g)$, and 
the lowest order part of (\ref{A4}) becomes
\begin{equation}
(2\beta_0+\gamma_i^{(0)}-\gamma_j^{(0)})S_{ij}
=\frac{\beta_1}{\beta_0}\gamma_i^{(0)}\delta_{ij}-G_{ij}
\label{A5}
\end{equation}
Dividing by $2\beta_0+\gamma_i^{(0)}-\gamma_j^{(0)}$ gives the expression
for $S_{ij}$ stated in Eq.~(\ref{eq:S:mat}), 
which was the one obtained previously by 
Buras {\em et al.} \cite{ref:buras:npb370,ref:buras:rmp52}. 

In the singular case where $2\beta_0+\gamma_i^{(0)}-\gamma_j^{(0)}$ vanishes,
(\ref{A4}) does not admit a consistent solution at lowest
order based on an expansion $J(g)=J+O(g)$. This case was left untreated
in the work of Buras {\em et al}. However, this case can be readily dealt with 
after realizing that $J(g)$ need not reduce to a constant matrix $J$ for
$g\to0$. In fact, all that is required is that $g^2J(g)\to0$ for $g\to0$, so 
that the expression (\ref{A1}) for $U(g,g_0)$ 
reduces to $U^{(0)}(g,g_0)$ for small
$g$ as it must. This allows for the possibility that $J(g)$ diverges for 
$g\to0$. Indeed, in the singular case (which can only occur when $i\ne j$)
(\ref{A4}) reduces at lowest order to
\begin{equation}
\beta_0\,g\,S_{ij}'(g)=-G_{ij}
\label{A6}
\end{equation}
which has the solution 
\begin{equation}
S_{ij}(g)=-\frac{G_{ij}}{\beta_0}\log(g)+c_{ij}
\label{A7}
\end{equation}
where $c_{ij}$ is an undetermined integration constant. 
In fact we do not need to 
determine $c_{ij}$ since it turns out not to contribute to the evolution
matrix at NLO. To see this, recall that the evolution matrix is determined at
NLO by $A(m_1,m_2)$ as in Eq.~(\ref{eq:U:4}), where now
\begin{eqnarray}
A(m_1,m_2)&=&\alpha_s(m_1)S(m_1)V^{-1}U^{(0)}(m_1,m_2)V
\nonumber \\
&&-\alpha_s(m_2)V^{-1}U^{(0)}(m_1,m_2)VS(m_2)\,. \nonumber \\
&&\label{A8}
\end{eqnarray}
Re-expressing (\ref{A7}) as 
\begin{equation}
S_{ij}(m)=-\frac{G_{ij}}{2\beta_0}\log(\alpha_s(m))+c_{ij}'
\label{A9}
\end{equation}
(where $c_{ij}'=c_{ij}-\frac{G_{ij}}{2\beta_0}\log(4\pi)$)
and substituting this into the expression for $A(m_1,m_2)_{ij}$ obtained from
(\ref{A8}) (recalling from Eq.'s (\ref{eq:U0})-(\ref{eq:a}) 
that $V^{-1}U^{0}(m_1,m_2)V$ is diagonal) 
we easily find that the constant $c_{ij}'$ drops out and our
previous expression Eq.~(\ref{eq:A:3}) is reproduced. 
This completes the alternative derivation of the NLO expression 
for $U(m_1,m_2)$. 
The argument also shows 
that for calculations involving $U(m_1,m_2)$ at NLO we may take $S_{ij}$ 
in the singular case to be given by (\ref{A9}) with $c_{ij}'\equiv 0$. 
This is useful for deriving the expressions Eq.'s 
(\ref{eq:tK4:1})--(\ref{eq:I2}) 
for the QED part of the evolution matrix at NLO.
%


\end{document}